\documentclass[twocolumn,superscriptaddress,nofootinbib,notitlepage,longbibliography,aps]{revtex4-1}
\pdfoutput=1

\usepackage{graphicx}
\usepackage{amsmath}
\usepackage{amssymb}
\usepackage{amsfonts}
\usepackage{physics}
\usepackage[dvipsnames]{xcolor}
\usepackage[linktoc=none]{hyperref}
\usepackage{braket}
\usepackage{tabstackengine}
\stackMath
\usepackage[mathscr]{euscript}
\usepackage{subcaption}
\usepackage{ragged2e}
\usepackage{soul}

\newcommand{\INFN}{INFN - Istituto Nazionale di Fisica Nucleare, sezione di Lecce, Italy}
\newcommand{\UNISA}{Dipartimento di Matematica e Fisica ''Ennio De Giorgi", Universit\`a del Salento, Italy}

\begin{document}

\title{ALP production from light primordial black holes: The role of superradiance}

\author{Marco Manno}
\email{marco.manno@unisalento.it}
\affiliation{\UNISA}
\affiliation{\INFN}

\author{Daniele Montanino}
\email{daniele.montanino@unisalento.it}
\affiliation{\UNISA}
\affiliation{\INFN}

\begin{abstract}
Light primordial black holes (LPBHs) with masses in the range $10$~g~$\leq M_{\rm BH} \leq 10^9$~g, although they evaporate before Big Bang Nucleosynthesis, can play a significant role in the production of both dark matter and dark radiation. In particular, LPBHs can evaporate into light axions or axion-like particles (ALPs) with masses $m_a \lesssim$~MeV, contributing to the effective number of neutrino species, $\Delta N_{\rm eff}$. Additionally, heavy scalar particles known as {\em moduli}, predicted by string theory, can be produced both via Hawking evaporation and through amplification by a mechanism called {\em superradiant instability} in the case of spinning primordial black holes (PBHs). These moduli can subsequently decay into ALPs, further amplifying their abundance.  
In this work, we calculate the number density of ALPs in the presence of moduli enhanced by superradiance for Kerr PBHs. Using current limits on $\Delta N_{\rm eff}$ from Planck satellite observations, we derive updated constraints on this scenario.
\end{abstract}
\maketitle

\section{Introduction}

Besides stellar-collapse black holes, primordial black holes (PBHs), produced by density fluctuations during post-inflationary eras, have been proposed for a long time \cite{Zeldovich:1967lct,Hawking:1971ei,Carr:1974nx,Escriva:2021aeh, Khlopov:2008qy}. Unlike stellar-collapse black holes, PBHs can, in principle, have a wide range of masses, from $\sim 1$~g to thousands of solar masses. For this reason, PBHs have been proposed as possible dark matter candidates \cite{Villanueva-Domingo:2021spv,Bird:2022wvk}. However, strong constraints on PBH masses and abundances exist due to various observations \cite{Green:2020jor,Green:2024bam} (updated bounds can also be found in \cite{PBHbounds}). Nevertheless, for PBHs in the mass range $10 \lesssim M_{\rm BH} \lesssim 10^9$~g, the constraints are much weaker \cite{Carr:2020gox, Papanikolaou:2020qtd,Domenech:2020ssp}. In the following, we refer to these PBHs as light primordial black holes (LPBHs). This is because LPBHs would evaporate via {\em Hawking radiation} \cite{Hawking:1975vcx} before neutrino decoupling, and their radiation products would enter into thermal equilibrium with the environment.\\ 
In addition to standard particles, LPBHs can also evaporate into non-standard heavy or light (or massless) particles. In the former case, if these particles are stable, they can serve as cold or warm dark matter candidates. This scenario has been extensively analyzed in the literature (see \cite{Fujita:2014hha,Lennon:2017tqq,Hooper:2019gtx,Baldes:2020nuv,Masina:2020xhk,Auffinger:2020afu,Gondolo:2020uqv,Masina:2021zpu,Cheek:2021odj,Cheek:2021cfe,Kim:2023ixo,Chen:2023lnj,Chen:2023tzd, Calza:2023iqa} for a non-comprehensive list). In the latter case, light particles could contribute to dark radiation. Essentially, there are two possible types of particles: gravitons \cite{Papanikolaou:2020qtd,Domenech:2020ssp,Arbey:2021ysg,Hooper:2020evu} or axion-like particles (ALPs) \cite{Schiavone:2021imu,Li:2022xqh,Agashe:2022phd}.

Besides evaporation, another mechanism that can enhance particle production from spinning black holes is {\em superradiance} (see \cite{Brito:2015oca} for a review). This is a phenomenon of amplification by coherent emitters, which is ubiquitous in many physical systems. For instance, waves scattering off a rotating cylinder can increase in amplitude. If a mirror surrounds the cylinder, the waves bounce back and forth, exponentially increasing their energy at the expense of the cylinder's kinetic energy. This phenomenon is called {\em superradiant instability}. In the case of spinning black holes, which can be produced in different scenarios \cite{He:2019cdb, Bai:2019zcd, Cotner:2019ykd, Harada:2017fjm, Eroshenko:2021sez, Flores:2021tmc}, the role of the scattering waves can be played by massive bosons, while the mirror is provided by gravity itself \cite{Press:1972zz,Teukolsky:1974yv,Cardoso:2004nk, Calza:2023rjt}. This leads to the formation of a {\em gravitational atom}, where a boson cloud is exponentially amplified. The key parameter governing this phenomenon is the {\em gravitational fine structure constant}, $\alpha = r_s / 2\lambda_c$, where $r_s = 2GM_{\rm BH}$ is the Schwarzschild radius of the black hole and $\lambda_c = 1 / m_S$ is the Compton wavelength of the boson mass \cite{Zouros:1979iw,Detweiler:1980uk,Dolan:2007mj,Witek:2012tr,Roy:2021uye,Bernal:2022oha,Bao:2022hew,Hoof:2024quk,Siemonsen:2022yyf}. Hereafter, we adopt natural units ($\hbar = c = k_B = 1$) and use $M_{\rm pl} = 1 / \sqrt{8\pi G}$. For $M_{\rm BH} < 10^9$~g, superradiance is effective for boson masses $m_S > 10^5$~GeV. If these particles are stable and feebly interacting with ordinary matter, they would not thermalize and could now constitute dark matter \cite{Bernal:2022oha}. Alternatively, if the massive bosons are unstable, they might decay into very weakly interacting light particles, such as axions or ALPs, leaving an excess of dark radiation beyond neutrinos.
String theory models often predict the existence of heavy moduli fields that behave like bosons. In such scenarios, moduli fields produced after inflation could decay into ALPs \cite{Cicoli:2012aq,Higaki:2012ar,Higaki:2013lra,Conlon:2013isa,Conlon:2013txa,Angus:2013sua,Evoli:2016zhj}. In addition, these particles can induce superradiant instabilities in primordial Black Holes, thereby enhancing the production of ALPs.
In this work, we analyze this scenario. In Sec.~\ref{Hawking}, we illustrate the main mechanism responsible for the mass and spin loss of Black Holes, namely Hawking radiation. In Sec.~\ref{superrad}, we present the phenomenon of superradiance and describe how it can increase the energy and spin loss of black holes. In Sec.~\ref{decay}, we introduce the decay mechanism of moduli into ALPs. In Sec.~\ref{complete}, we describe the complete scenario, incorporating both superradiance and decay, and show the evolution of black hole mass and spin, as well as the comoving number density of moduli and ALPs for specific parameter values, highlighting the contribution of superradiance. In Sec.~\ref{extra}, we compute the contribution of additional radiation due to ALPs and compare it with current bounds from Planck. Finally, we present our conclusions in Sec.~\ref{conclusions}.

\section{Hawking radiation}\label{Hawking}

Hawking radiation, a theoretical prediction made by physicist Stephen Hawking
in 1974 \cite{Hawking:1974rv,Hawking:1975vcx}, is the process by which black holes can emit radiation. Here we focus on Kerr black hole. In Boyer-Lindquist coordinates, a Kerr black hole is described by the line element
\begin{align}
	ds^2=&-\frac{\Delta-a^2\sin^2\theta}{\Sigma}dt^2+\Sigma\left(\frac{dr^2}{\Delta}+ d\theta^2\right) \notag \\&+\frac{(r^2+a^2)^2 -a^2\Delta\sin^2\theta}{\Sigma}d\varphi^2
    -\frac{2r_s ra\sin^2\theta}{\Sigma}dtd\phi\,,
	\label{lineelement}
\end{align}
where $a = J/M_{\rm BH}$ and
\begin{equation}
	\Delta = r^2+a^2-r_s r\;, \quad\quad \Sigma=r^2+a^2\cos^2\theta~.
\end{equation}
In this geometry we have the true singularity at $r=0$ and two singularities at $r_{\pm}=r_s/2\pm\sqrt{(r_s/2)^2-a^2}$.  The angular velocity of the black hole at the event horizon is given by
\begin{equation}
	\Omega \equiv \frac{a}{r_+^2+a^2} .
\end{equation}
Unlike Schwarzschild black hole, which is non-rotating and spherically symmetric, a Kerr black hole possesses angular momentum. This directly influences the Hawking temperature
\begin{equation}
     T_{\rm BH} = \frac{1}{4\pi G M_{\rm BH}}\frac{\sqrt{1-a_\star^2}}{1+\sqrt{1-a_\star^2}}\,,
\end{equation}
in which $a_{\star} = 2a/r_s=J/(GM_{\rm BH}^2)$.

The spectra of emitted particles for a particle species $i$ \cite{Page:1976df,Dong:2015yjs} is:
\begin{equation}
    \frac{d^2 \mathcal{N}_{i}}{d E d t}=\sum_{\mu}\frac{\Gamma_{s_i}^{\mu}(M_{\rm BH},a_\star,E)/2\pi}{\exp\left[(E - \mu\Omega)/T_{\rm BH}\right]-( -1)^{2s_i}}\,,
    \label{phasespaceint}
\end{equation}
with $\Gamma_{s_i}^{\mu}(M_{\rm BH},a_\star,E)=\sum_l\sigma_{s_i}^{l\mu}(M_{\rm BH},a_\star,E)(E^2-m_i^2)/\pi$, where $\sigma_{s_i}^{l\mu}$ the absorption cross-section of the particle  with total (axial) angular momentum $l$ ($\mu$) \cite{MacGibbon:1990zk}. Only particles with mass $m_i<T_{\rm BH}$ can be emitted by the black hole. In contrast with the Schwarzschild case where only the mass is reduced during the evaporation process, radiating particles carry away angular momentum causing the spin loss of the black hole \cite{Cheek:2022mmy}
\begin{align}
 \frac{dM_{\rm BH}}{dt} &= - \varepsilon(M_{\rm BH}, a_\star)\frac{M_{\rm pl}^4}{M_{\rm BH}^2}\,, \label{masshawk} \\
 \frac{da_\star}{dt} &= - a_\star[\gamma(M_{\rm BH},a_\star) - 2\varepsilon(M_{\rm BH},a_\star)]\frac{M_{\rm pl}^4}{M_{\rm BH}^3}\,, \label{spinhawk}
\end{align}
where $\varepsilon(M_{\rm BH}, a_\star)=\sum_i  \varepsilon_i(M_{\rm BH}, a_\star)$, $\gamma(M_{\rm BH},a_\star) =\sum_i \gamma_i(M_{\rm BH},a_\star)$ with
\begin{align}
        \varepsilon_i(M_{\rm BH}, a_\star) &= \frac{1}{2\pi}\int_{\zeta_i}^\infty\sum_{\mu}\frac{\xi\Gamma_{s_i}^{\mu}(M_{\rm BH},a_\star,\xi)}{e^{\xi^\prime}-(-1)^{2s_i}}d{\xi} \;,    \label{eps}        \\ 
        \gamma_i(M_{\rm BH},a_\star) &= \frac{4}{ a_\star}\int_{\zeta_i}^\infty\sum_{\mu}\frac{\mu\Gamma_{s_i}^{\mu}(M_{\rm BH},a_\star,\xi)}{e^{\xi^\prime}-(-1)^{2s_i}}d{\xi}\;,\label{gamma}
\end{align}
being $\xi = EM_{\rm BH}/M_{\rm pl}^2$, $\zeta_i =m_iM_{\rm BH}/M_{\rm pl}^2$, and
\begin{align*}
    \xi^\prime = \frac{\xi - \mu\Omega(a_\star) M_{\rm BH}/M_{\rm pl}^2}{2}\left(1+\frac{1}{\sqrt{1-a_\star^2}}\right)\,,
\end{align*}
in order to have a smooth transition from Kerr to Schwarzschild in the limit $a_\star \rightarrow 0$. These functions depend mildly on the parameters $M_{\rm BH}$ and $a_\star$  \cite{Page:1976df}.

\section{Superradiance}\label{superrad}

In addition to Hawking radiation, rotating black holes exhibit another remarkable phenomenon: superradiance, which can lead to significant energy
extraction and even instabilities. Superradiant instabilities occur when the conditions for wave amplification lead
to a runaway effect. 
The instabilities can be triggered in various ways (see \cite{Brito:2015oca}), but here we consider the more natural mechanism called \textit{gravitational atom}, i.e. the system of a black hole and a massive bosonic field.
Massive scalar fields with mass $m_S$ can be trapped in hydrogen-like bound states:
\begin{equation}
  E_n = \omega_n \simeq m_S - \frac{\alpha^ 2 m_S}{2 n^2}      \;,
\end{equation}
where we have defined a useful parameter called gravitational coupling:
\begin{equation}
    \alpha = \frac{r_s}{2\lambda_c} = G M_{\rm BH} m_S \simeq 0.38 \left(\frac{M_{\rm BH}}{10^7 g}\right) \left(\frac{m_S}{10^7 \text{GeV}}\right) \;, \label{gravcoupling}
\end{equation}
in which $\lambda_c = 1/m_S$ is the Compton length of the boson.
For superradiance to occur, the wave frequency $\omega$ must satisfy a specific condition relative to the black hole's spin. This condition is expressed as:
\begin{equation}
\omega < \mu\Omega\; ,
\label{eq:superradiance_condition}
\end{equation}
where $\mu$ is the azimuthal quantum number of the field.
When this condition is met, an incident wave is amplified, extracting rotational energy from the black hole. This amplification leads to the growth of bound modes in the case of massive bosonic fields. The growth rate can be approximated as \cite{Bernal:2022oha,Detweiler:1980uk} 
\begin{equation}
    \Gamma_{\text{sr}}(M_{\rm BH},a_\star) \simeq \frac{m_S}{24} \left(\frac{m_S M_{\rm BH}}{8 \pi M_{\rm pl}^2}\right)^8 (a_\star -2 m_S r_+)\;, \label{growthrate}
\end{equation}
where the dominant unstable mode $n=2$ and $l=\mu=1$ was considered.

We notice that $\Gamma_{sr}$ is calculated at order ${\cal O}(\alpha)$ in the limit of $\alpha \ll 1$, i.e., the non relativistic regime. Corrections are calculated in the literature \cite{Hoof:2024quk} up to ${\cal O}(\alpha^5)$ or even numerically \cite{Siemonsen:2022yyf}. However, the approximated Eq.~(\ref{growthrate}) is enough to our purposes.
It's straightforward now to generalize Eqs.~(\ref{masshawk}) and (\ref{spinhawk}) in the presence of superradiance. \\
Considering the fact that superradiant instabilities can amplify the scalar field, we can indicate with ${\cal N}_S$ the number of scalar particles gravitationally bounded to a PBH. For nearly stable scalar particles, the dynamics of the system can be described as follows \cite{Bernal:2022oha}:
\begin{align}
   \frac{d{\cal N}_S}{dt} &= \Gamma_{sr}(M_{\rm BH}, a_\star) {\cal N}_S\;, \label{numbermoduli}\\
    \frac{dM_{\rm BH}}{dt}&= - \varepsilon(M_{\rm BH}, a_\star)\frac{M_{\rm pl}^4}{M_{\rm BH}^2}- m_S \frac{d{\cal N}_S}{dt} \;, \label{masssuper}\\
    \frac{da_{\star}}{dt}&= - a_\star[\gamma(M_{\rm BH},a_\star) - 2\varepsilon(M_{\rm BH},a_\star)]\frac{M_{\rm pl}^4}{M_{\rm BH}^3}+\notag\\&-\frac{1}{G M_{\rm BH}^2} \left(\sqrt{2} - 2 \alpha a_{\star}\right) \frac{d{\cal N}_S}{dt} \;, \label{spinsuper}
\end{align}
where the factor $\sqrt{2}=\sqrt{l(l+1)}$ in the last term the angular momentum carried away by a single boson in the dominant mode $l=1$. It has been shown that the initial number of scalar particles does not affect the final result, so one can set ${\cal N}_S(0)=1$.
As we will see later, the superradiant phase is much faster compared to the evaporation timescale. For this reason, one can, in principle, neglect evaporation during the superradiant phase, although no approximations are made in our calculations.
We highlight that numerical simulations indicate that for $a_{\star} \simeq 1$, the superradiant instability reaches its maximum at $\alpha \sim 0.42$ \cite{Dolan:2007mj,Witek:2012tr}.

\begin{figure*}[tbh!] \centering \includegraphics[width=0.9\linewidth]{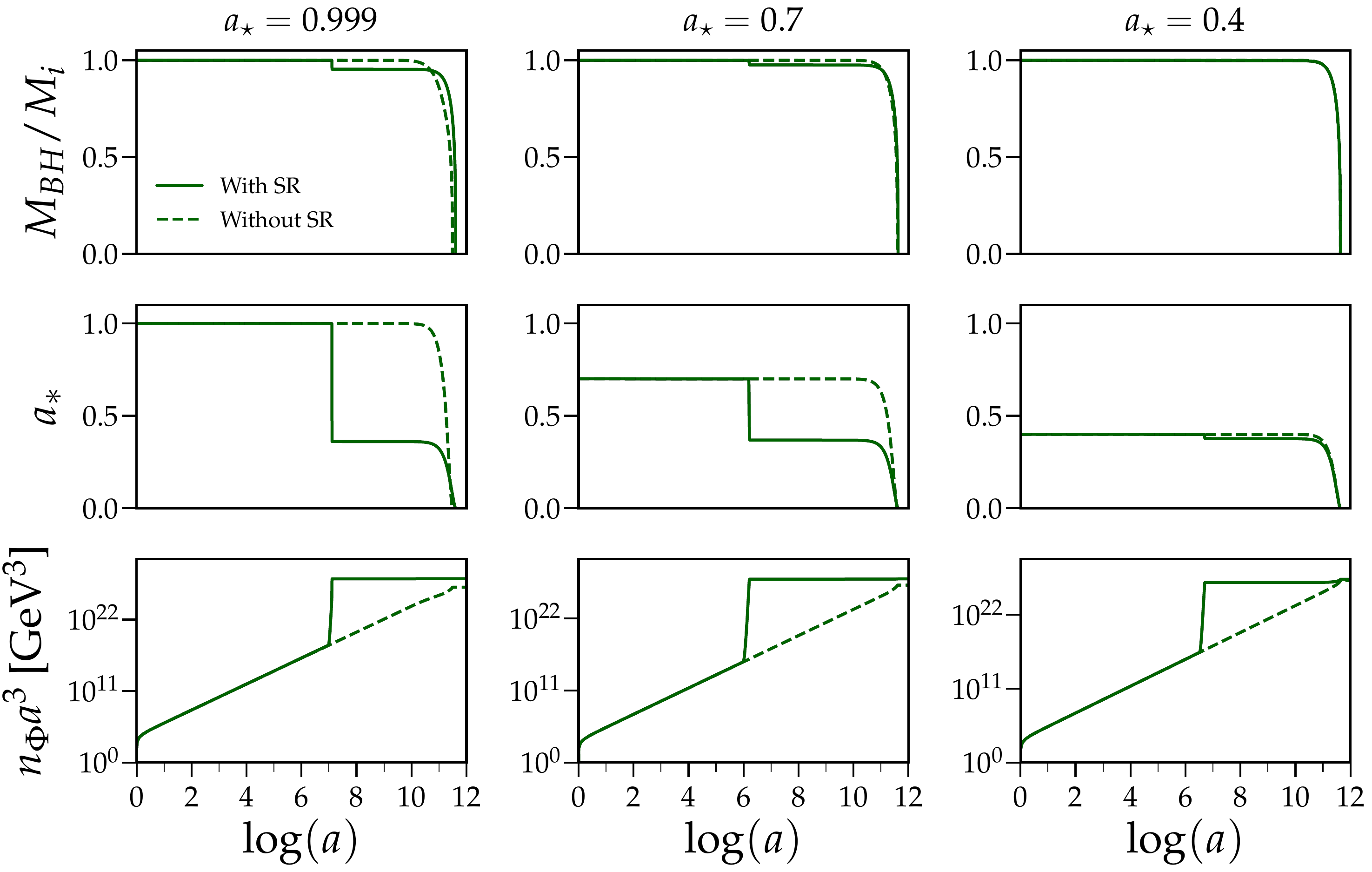} \caption{\justifying Interplay between Hawking radiation and superradiance for different initial spins $a_\star$. The graph shows the evolution of the mass (upper row), the spin (middle row), and the comoving number of moduli $n_{\Phi} a^3$ (bottom row) as a function of (the decimal log of) the cosmological scale factor $a$. Solid lines represent the evolution with superradiance, while dashed lines represent the evolution without it. The parameters used are $M_i = 2.6 \times 10^6$~g , $m_\Phi=10^7$~GeV and $\Omega_{{\rm PBH},i} = 10^{-15}$.}
\label{firstplots} \end{figure*}

\section{ALP production from moduli decay}\label{decay}

In this work, we consider the heavy particles called moduli, $\Phi$. These fields are real scalar fields predicted by string theory, with couplings suppressed by the Planck constant $M_{\rm pl}$, and naturally arise from the complex Calabi-Yau geometry (see \cite{McAllister:2023vgy} for a recent review). These are naturally long-lived particles because their coupling with matter particles is suppressed by $1/M_{\rm pl}$. The moduli $\Phi$ can serve as the scalar particles responsible for inducing superradiance in PBHs. From Eq.~(\ref{gravcoupling}), we observe that for a moduli mass $10^5\lesssim m_\Phi\lesssim 10^{13}$ GeV, superradiance is effective, which aligns with the Large Volume Compactification scenario \cite{Cicoli:2012aq,Higaki:2012ar,Higaki:2013lra,Conlon:2013isa,Conlon:2013txa,Angus:2013sua,Evoli:2016zhj}. For simplicity, we assume that no other heavy scalar particles contribute to the superradiant instability besides moduli.
As we will demonstrate in the following section (see also \cite{Bernal:2022oha}), the timescale of moduli production via superradiance is much shorter than the moduli decay lifetime, $\tau_{sr}\ll\tau_\Phi$.

A population of ALPs can be produced by the decay of moduli particles. The decay rate of moduli $\Phi$ is \cite{Cicoli:2012aq}:
\begin{equation}
    \Gamma_{\Phi} =\tau_\Phi^{-1}= \frac{1}{4 \pi} \frac{m_{\Phi}^3}{(M_{\rm pl} /k)^2} \;,
\end{equation}
with $k$ a constant of order ${\cal O}(1)$  (in the following we choose for simplicity $k=1$). 
Since moduli have gravitational couplings, they can decay into various sectors, including both standard and non-standard particles. Without constraints on the mass of the moduli, the universe could remain dominated by these scalar particles, delaying the onset of primordial nucleosynthesis (BBN). To align with BBN predictions, reheating must occur before nucleosynthesis, requiring a reheating temperature $T_{\text{reheat}} > {\cal O}(1) \, \text{MeV}$, which imposes an upper limit on the moduli mass, $m_{\Phi} \gtrsim 30 \, \text{TeV}$ \cite{Cicoli:2012aq}. Moduli do not decay instantly and the energies of the produced ALPs scale differently depending on when the moduli decayed. ALPs are produced by processes
$\Phi \rightarrow aa$ with $\Gamma_{\Phi \rightarrow aa} = B_a \Gamma_{\Phi}$, 
where $B_a$ is the branching ratio, which quantifies how many moduli decay in ALPs. Equivalently $1-B_a$ quantifies how many moduli decay in standard model particles.

\section{Complete scenario}\label{complete}

Taking into account all the production channels, the evolution equations for the energy densities can be cast as follows 
\begin{align}
   \dot{\rho}_{\rm SM} + 4 H \rho_{\rm SM} &= \varepsilon_{\rm SM}\frac{M_{\rm pl}^4}{M_{\rm BH}^2} n_{\rm BH}+\Gamma_{\Phi} (1-B_a) \rho_{\Phi}\;, \\
   \dot{\rho}_{a} + 4 H \rho_{a} &= \varepsilon_a\frac{M_{\rm pl}^4}{M_{\rm BH}^2} n_{\rm BH} +\Gamma_{\Phi} B_a \rho_{\Phi}\;,
\end{align}
where $\rho_\Phi=m_\Phi n_\Phi$, where $n_\Phi=n_{\rm BH}{\cal N}_\Phi$ is the number density of moduli and $n_{\rm BH}(t)=n_{{\rm BH},i}/a^3(t)$ is the number density of PBHs. $\varepsilon_{\rm SM}$ is the sum of the $\varepsilon_i$ in Eq.~(\ref{eps}) extended on standard model particles and $\varepsilon_{a}$ for the ALP field. 

Similarly, we can write the equation for $\rho_\Phi$
\begin{equation}
    \dot{\rho}_{\Phi} + 3 H \rho_{\Phi} =  \Gamma_{sr}\rho_\Phi+\varepsilon_{\Phi}\frac{M_{\rm pl}^4}{M_{\rm BH}^2}n_{\rm BH}-\Gamma_\Phi \rho_\Phi\;,
\end{equation}
where the first term takes into account the superradiant production, the second term the evaporation and the last one the decay. Previous equations must be supplemented by the Friedmann-Robertson-Walker equation:
\begin{equation}
    H^2(t) = \left(\frac{\dot{a}}{a}\right)^2 = \frac{8 \pi G}{3} \left(\rho_{\rm SM}+ \rho_a+
 \rho_\Phi + \rho_{\rm BH}\right)\;,
\end{equation}
with $\rho_{\rm BH}(t)=M_{\rm BH}(t)n_{\rm BH}(t)$.
This equation allows for solving the evolution equation in terms of the cosmological scale factor $a$ instead of the time $t$:
\begin{equation}
   \frac{d}{dt} = \frac{da}{dt}\cdot\frac{d}{da}  = a H(a) \cdot\frac{d}{da}\;,
\end{equation}
with $a=1$ the scale factor at $t=t_{i}$ when PBH is formed.  This time roughly corresponds to the time when
the Schwarzschild radius of the perturbation was of the order of the horizon scale, $2GM_{\rm BH}\sim 2t_i$ \cite{Zeldovich:1967lct}.

In order to simulate the entire system, we utilized the public code \texttt{FRISBHEE} \cite{Cheek:2021odj, Cheek:2021cfe, Cheek:2022dbx, Cheek:2022mmy}. However, we extended the code to include the effects of superradiant instabilities and the decay of moduli particles. The initial condition is the temperature of the universe at the time of PBH formation. Since we choose $\Omega_{{\rm PBH},i}\ll 1$, the universe is radiation dominated at formation epoch. This corresponds to a temperature 
\begin{equation}
    T_i = \left(\frac{90}{32 \pi^3  Gg_\rho(T_i)t_i^2}\right)^{1/4}\;, \label{temp}
\end{equation}
where $g_\rho(T)$ is the number of degrees of freedom for energy density at temperature $T$. For simplicity we consider only the degrees of freedom  of the standard model at high temperature ($T_i>$~TeV), that is $g_\rho=106.75$ \cite{Husdal:2016haj}.

Furthermore, we note that for our range of masses, as pointed out in \cite{Bernal:2022oha}, the PBHs do not evaporate before the superradiant instability becomes relevant.
\begin{figure*}[tbh!]
    \centering
\includegraphics[width=0.8\linewidth]{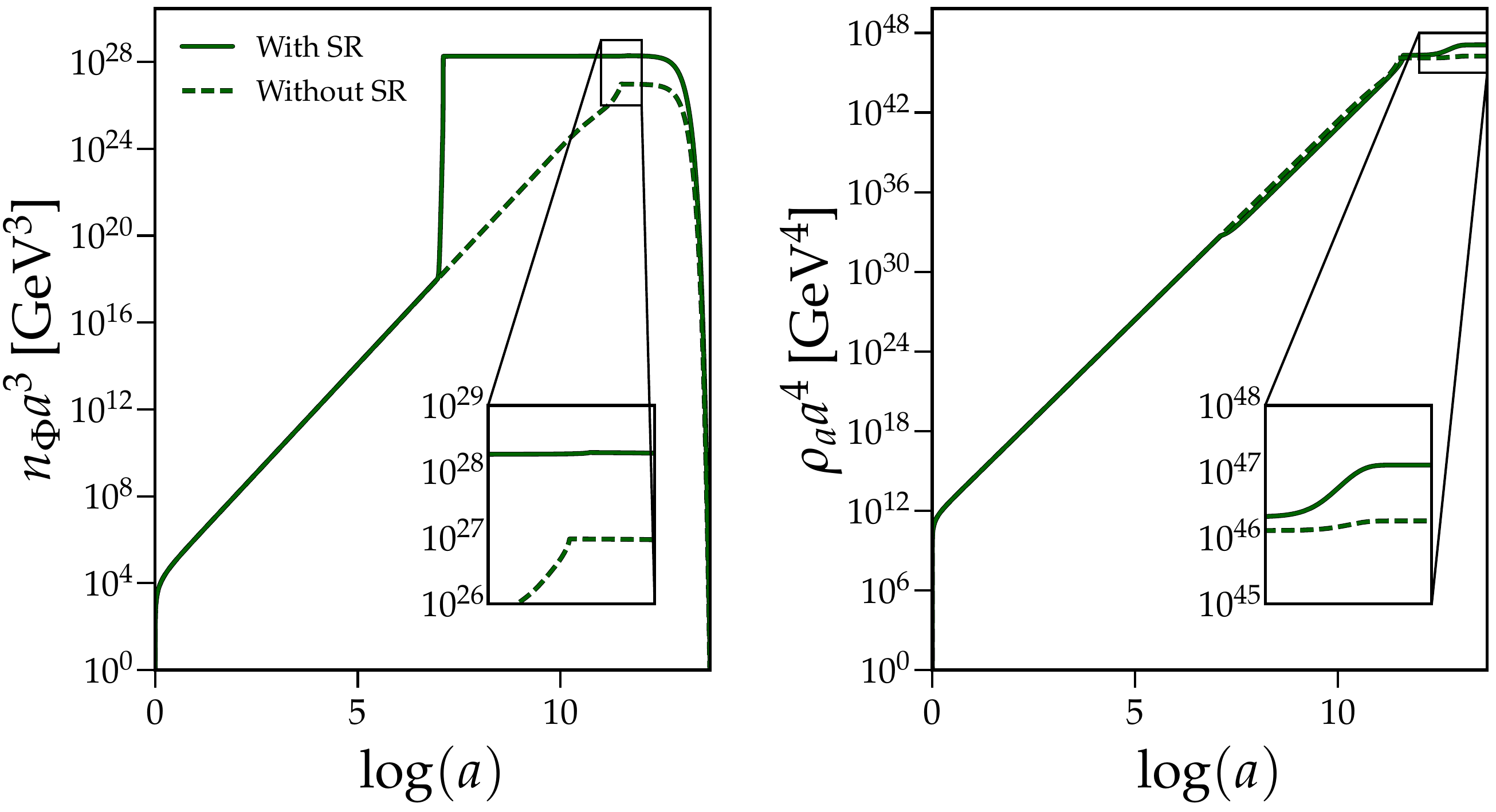}
    \caption{\justifying   Comoving number of moduli $n_{\Phi} a^3$ (left) and comoving axion radiation $\rho_a a^4$ (right) as a function of the scale factor $a$. Superradiance amplifies the number of moduli $n_{\Phi} a^3$, leading to an increase in the radiation $\rho_a$. The values used are $M_i = 2.6 \times 10^6$ g, $m_{\Phi}=10^7 \, \mathrm{GeV}$, $\Omega_{{\rm PBH},i} = 10^{-15}$ and $B_a =0.1$.}
\label{modplot}
\end{figure*}

The plots presented in Fig.~\ref{firstplots} illustrate the scenario in which moduli decay is temporarily not considered. This choice is made to better highlight the interplay between Hawking radiation and superradiance. However, the effects of moduli decay are fully taken into account in subsequent figures. In the first column, the PBH has an initial spin of $a_{\star} = 0.999$, in the second column the spin is reduced to $a_{\star} = 0.7$, and in the third one the spin is further decreased to $a_{\star} = 0.4$. The initial mass of the PBH is $M_i = 2.6 \times 10^6$ g, the mass of the moduli is $10^7 \, \mathrm{GeV}$, and the initial fraction of the PBH energy density $\Omega_{{\rm PBH},i}=M_in_{{\rm BH},i}/\rho_i$ is set to $\Omega_{{\rm PBH},i} = 10^{-15}$. These parameters are used as illustrative examples. However, the relationship between the mass of the PBH and the moduli is critical, as it ensures that the gravitational coupling $\alpha$ remains of the order ${\cal O}(0.1)$, which is necessary for efficient superradiant amplification.

A remark is in order. The initial mass distribution depends on the formation mechanism  \cite{Tashiro:2008sf,Carr:2017jsz,Germani:2018jgr}. However, in order to not increase the number of parameters, in this paper we have considered a single value for initial primordial black hole mass and spin. The mass evolution of the PBHs, shown in the first row, reflects the competing effects of Hawking radiation and superradiance. Superradiance plays a progressively smaller role as the initial spin decreases. The mass evolution is mainly driven by radiation at $t \lesssim t_{\text{ev}}$, although we observe a small decrease at $t \sim t_{sr} \sim 1/\Gamma_{sr}$.
The second row depicts the evolution of the PBH dimensionless spin parameter $a_{\star}$. For the highest spin case ($a_{\star} = 0.999$), superradiance rapidly depletes the spin. This dramatic loss of angular momentum is a clear signature of the efficiency of superradiance when the initial spin is high. In the second scenario ($a_{\star} = 0.7$), the spin also decreases, but at a slower rate. Finally, for the lowest spin case ($a_{\star} = 0.4$), the spin evolution is much less affected by superradiance, and the reduction in spin is minimal compared to the first two cases.
The third row, which shows the comoving number density of moduli $n_{\Phi} a^3$, provides the most striking result. In the case of the highest initial spin, the amplification due to superradiance is significant. After the superradiance ends, Hawking radiation continues, but its contribution is negligible compared to the abundance produced by superradiance.
As we reduce the initial spin, the difference between the production of moduli with and without superradiance is much smaller. 

In Fig.~\ref{modplot} we present a simulation for the total set of equations including the moduli decay, for $M_i = 2.6 \times 10^6$ g, $m_{\Phi}=10^7 \text{GeV}$, $\Omega_{{\rm PBH},i} = 10^{-15}$ and $B_a =0.1$.
The left panel illustrates the evolution of the number of moduli particles $N_{\Phi}$ as a function of the logarithmic scale factor $\log(a)$. We notice that the number of moduli grows by a factor of approximately ${\cal O}(10^{10})$ during the superradiant phase. 
The plot on the right-hand side of the graph illustrates the comoving axion energy density, \( \rho_a a^4 \). Two key features of this plot need to be justified: the small decrease between the solid and dashed lines and the final increase. The small decrease observed corresponds to a point where the black hole (BH) spin decreases significantly, as shown in Fig.~\ref{firstplots}. This decrease in spin leads to a reduction in the evaporation factor, \( \epsilon_a \). Consequently, the comoving axion density decreases as a direct result. This behavior is consistent with findings in the work \cite{Taylor:1998dk}, which demonstrates that \( \epsilon_a \) diminishes when the spin of the black hole decreases. The final increase in the comoving axion energy density is due to the decay of moduli (enhanced by superradiance) into axions. This effect becomes dominant and leads to a higher axion density compared to the case which only accounts for Hawking radiation contributions.

\section{Extra effective number of neutrinos}\label{extra}

\begin{figure*}
    \centering
\includegraphics[width=0.49\linewidth]{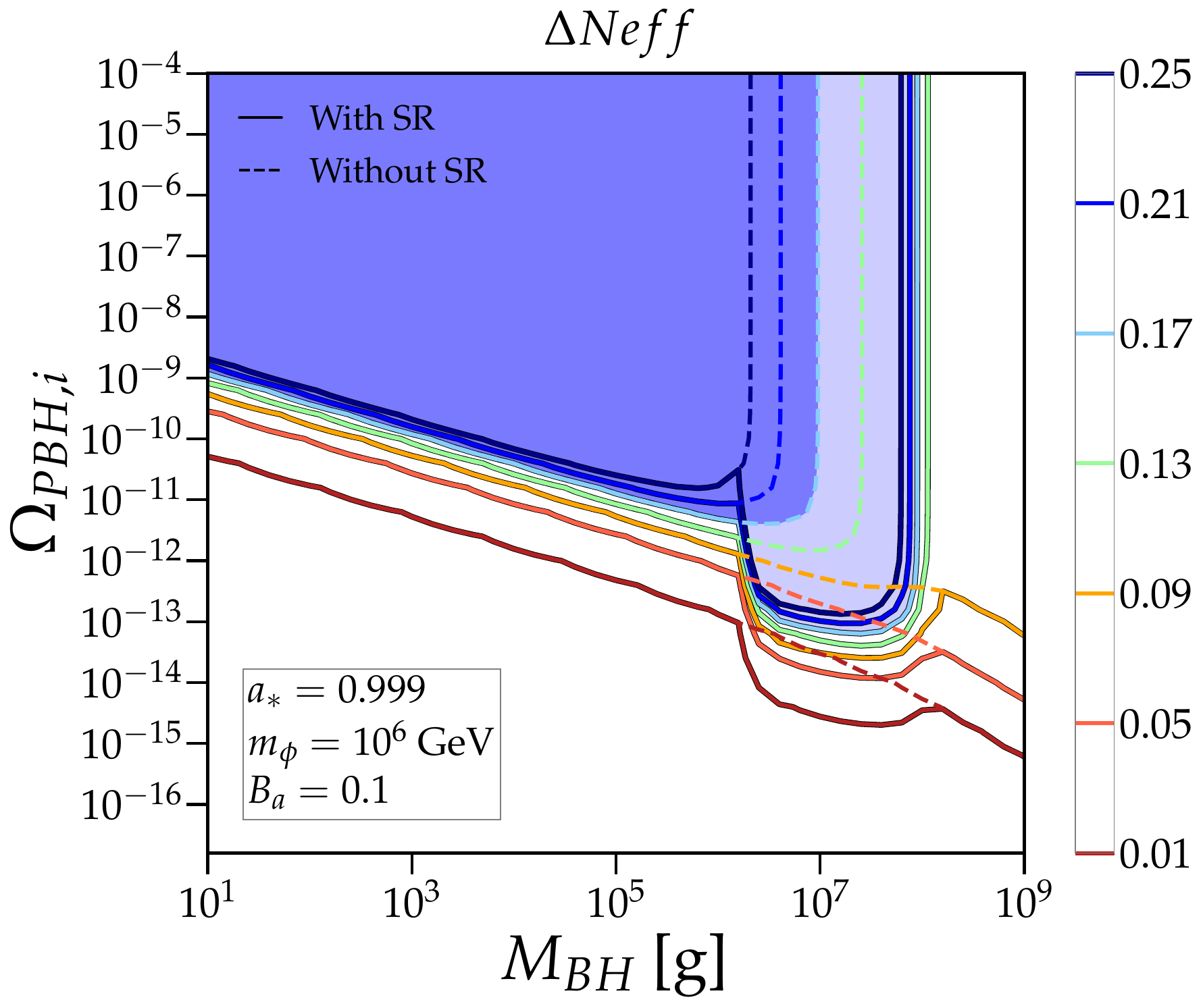}
    \includegraphics[width =0.49\linewidth]{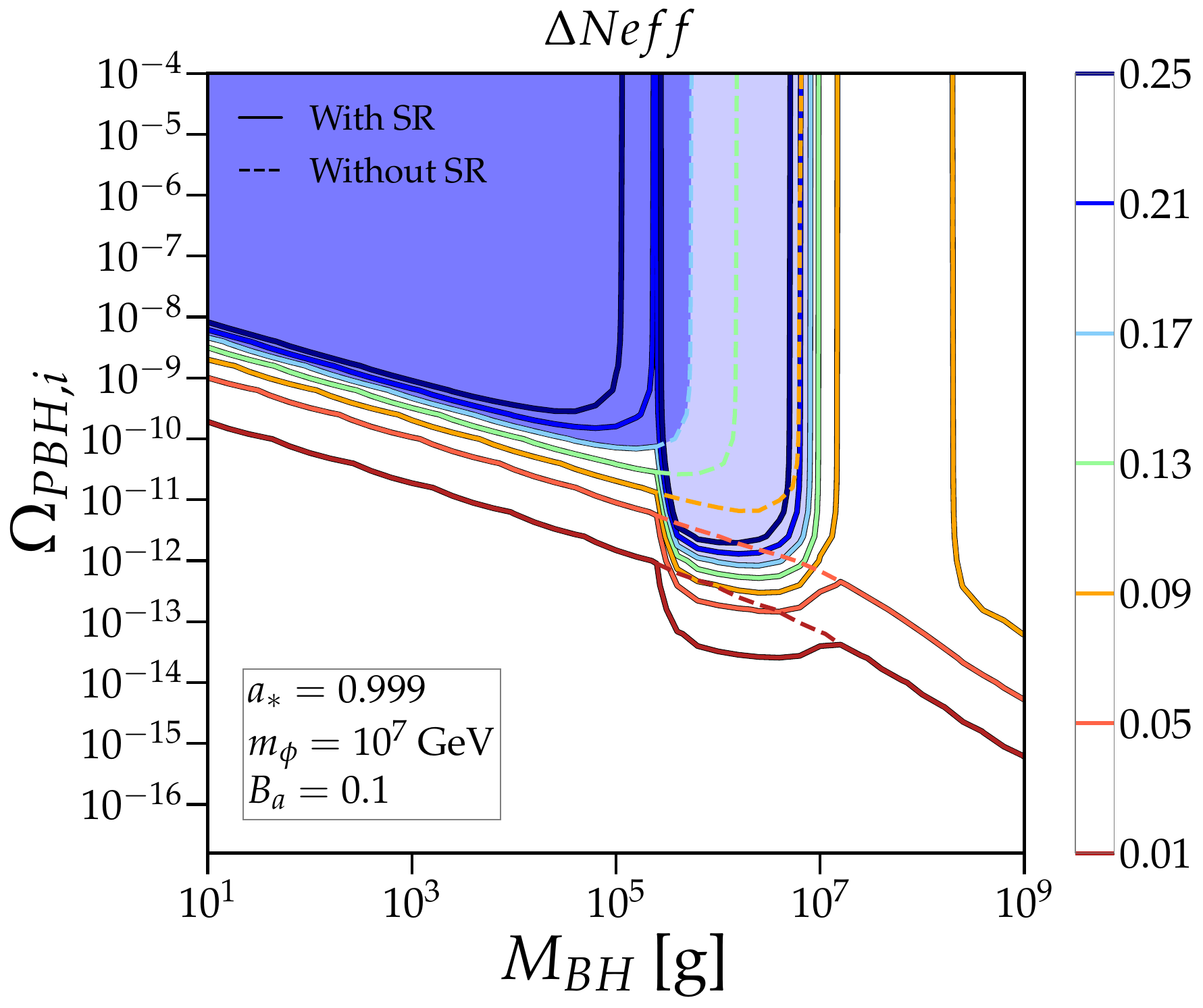}
    \caption{\justifying Contour lines representing the extra effective number of neutrinos in the case where superradiance is considered (solid), and without superradiance (dashed). The shaded areas represent the Planck limits. The left panel considers $m_\Phi = 10^6 \, \mathrm{GeV}$, while the right panel assumes $m_\Phi = 10^7 \, \mathrm{GeV}$. The superradiant scenario significantly extends the range of $\Delta N_{\rm eff}$ values.
}
    \label{neff}
\end{figure*}
The inclusion of superradiance not only amplifies the production of moduli
but also leaves a lasting impact on the axion population, which have observational
consequences for the Cosmological Axion Background (CAB) \cite{Conlon:2013isa}. This axion population can contribute to the extra effective number of neutrino species:
\begin{equation}
    \Delta N_{\rm eff} = \frac{\rho_a(T_0)}{\rho_{r}(T_0)} \left[N_{\rm eff} + \frac{8}{7} \left(\frac{11}{4}\right)^{4/3}\right]\;, \label{effectivenumbertoday}
\end{equation}
where $T_0=2.73$~K is the current temperature of the cosmic microwave background (CMB), $N_{\rm eff}=3.043$ is the effective number of neutrinos \cite{Cielo:2023bqp}, and $\rho_{r}$ denotes the "standard" radiation made of photons and neutrinos. It has been pointed out that the impact of dark radiation on $\Delta N_{\rm eff}$ may help alleviate the Hubble tension \cite{Papanikolaou:2023oxq}.
Now we want to redshift the latter equation to the time at which the decay of the moduli particles is $t_d\gtrsim  \Gamma_\Phi^{-1}$. We denote this time as $t_d$, and the temperature at this time as $T_d$.

As the universe expands the standard radiation energy density gets diluted as
\begin{equation}
 \frac{\rho_r(T_0)}{\rho_{\rm SM}(T_d)} = \frac{g_{\rho}(T_0)}{g_{\rho}(T_d)} \left(\frac{T_0}{T_d}\right)^4 \;.
\end{equation}
Further, total entropy must be conserved, so we have
\begin{equation}
    \frac{T_0}{T_d} = \frac{a(T_d)}{a(T_0)} \left(\frac{g_S(T_d)}{g_S(T_0)}\right)^{1/3}\;,
\end{equation}
where $g_S$ are the entropy degree of freedom, and thus
\begin{equation}
    \frac{\rho_r (T_0)}{\rho_{\rm SM}(T_d)} =   \frac{g_{\rho}(T_0)}{g_{\rho}(T_d)}\left(\frac{g_S(T_d)}{g_S(T_0)}\right)^{4/3}\left(\frac{a(T_d)}{a(T_0)}\right)^4\;.
\end{equation}
Differently, the dark radiation density just gets diluted by the expansion of the universe:
\begin{equation}
    \frac{\rho_a(T_0)}{\rho_a(T_d)} = \left(\frac{a(T_d)}{a(T_0)}\right)^4\;.
\end{equation}
Then the ratio between the present energy densities is expressed in terms of the energy densities at the time of decay as
\begin{equation}
    \frac{\rho_a(T_0)}{\rho_r(T_0)} = \frac{g_{\rho}(T_d)}{g_{\rho}(T_0)} \left(\frac{g_S(T_0)}{g_S(T_d)}\right)^{4/3}\frac{\rho_a(T_d)}{\rho_{\rm SM}(T_d)} \;. 
\end{equation}
Inserting the latter in Eq.~(\ref{effectivenumbertoday}) we get the extra effective number of neutrinos:
\begin{equation}
        \Delta N_{\rm eff} = 7.446\times\frac{g_{\rho}(T_d)}{g_{\rho}(T_0)} \left(\frac{g_S(T_0)}{g_S(T_d)}\right)^{4/3} \frac{\rho_a(T_d)}{\rho_{\rm SM}(T_d)} \;.
\end{equation}

\begin{figure}[tb!]
    \centering
    \includegraphics[width=1.\linewidth]{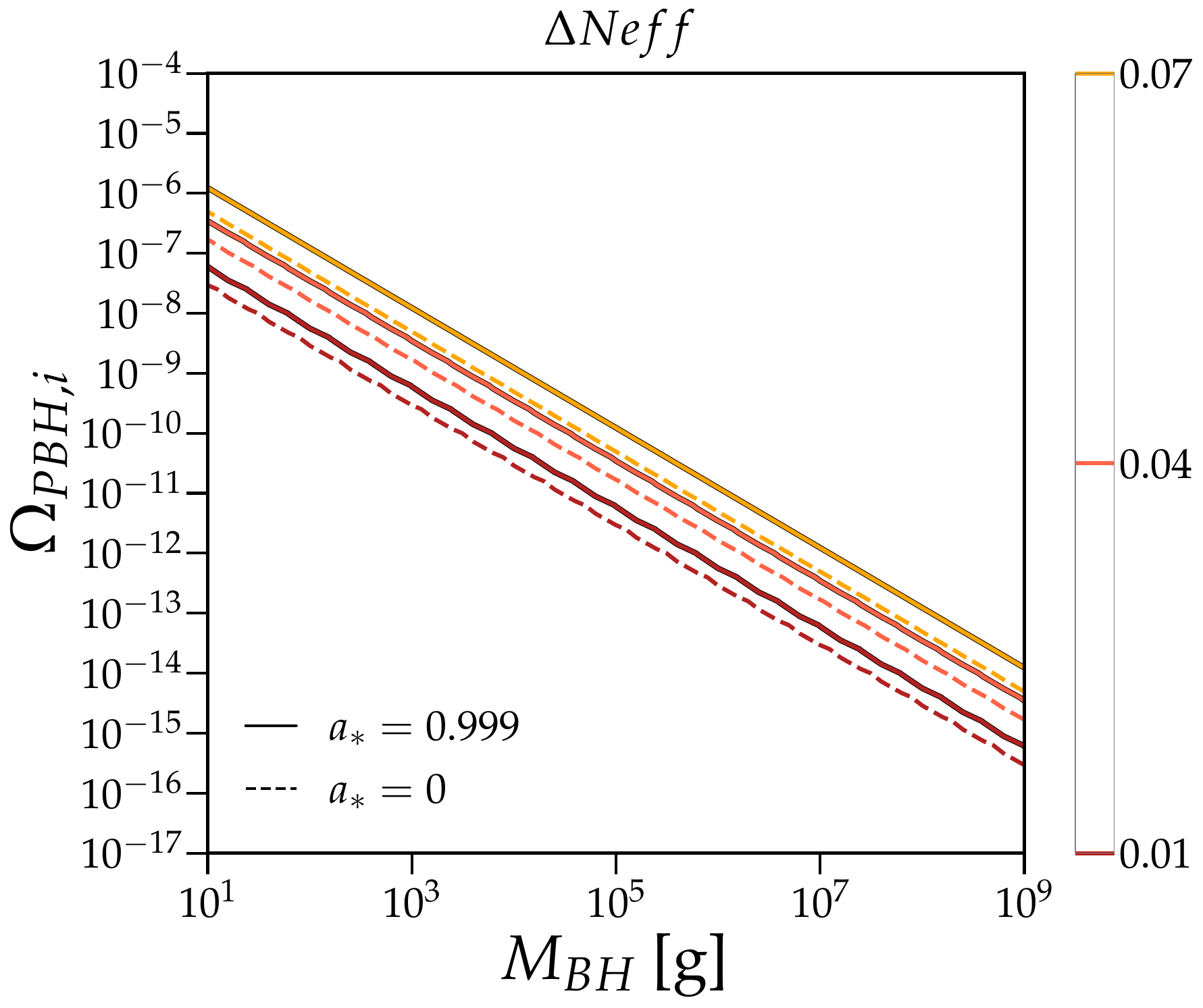}
\caption{\justifying  Extra effective number of neutrinos without the presence of moduli. The production of ALPs is driven solely by Hawking radiation in two possible cases: $a_\star = 0$ (dashed) and $a_\star = 0.999$ (solid). $\Delta N_{\rm eff}$ remains relatively low across the parameter space, compared to the previous image.}
   \label{neffnomod}
\end{figure}

Below we present the main results of our work. Figure ~\ref{neff} showcases contour plots of the extra effective number of neutrinos, $\Delta N_{\rm eff}$, as a function of the PBH mass on the x-axis and the initial PBH abundance relative to radiation, $\Omega_{{\rm PBH},i}$, on the y-axis. These results provide insight into how $\Delta N_{\rm eff}$ evolves across different parameter spaces in two distinct scenarios: with and without the inclusion of the superradiance effect.\\
The left panel of Fig.~\ref{neff} shows the results obtained with a moduli mass of $m_{\Phi} = 10^6 \, \text{GeV}$. In this simulation, the PBH spin is fixed at $a_\star = 0.999$, and $B_a= 0.1$. For this value of mass, Eq.~(\ref{temp}) yields a decay temperature $T_d\simeq 100$~MeV, which corresponds to $g_S\simeq g_\rho\simeq 18$ \cite{Husdal:2016haj}.
The right panel of Fig.~\ref{neff} is the same as the left panel, but with a moduli mass of $m_{\Phi} = 10^7 \, \text{GeV}$. As expected, increasing the moduli mass shifts the effect of superradiance to lower values of $M_{\rm BH}$. The solid lines represent the case where superradiance is taken into account, while the dashed lines correspond to the case without superradiance. A key observation is that superradiance significantly extends the range of $\Delta N_{\rm eff}$ values. This is visually evident from the fact that the contour lines shift downwards, indicating that for a given PBH mass and initial abundance, the inclusion of superradiance leads to higher values of $\Delta N_{\rm eff}$. The enhancement caused by superradiance is more visible when $\alpha \sim {\cal O}(0.1)$, as we expect.
Furthermore, the shaded regions indicate the observational limits derived from Planck data, i.e., $\Delta N_{\rm eff}<0.17$ at 68\% C.L.\ \cite{Planck:2018vyg}. 
The dark blue shaded region corresponds to the exclusion limits in the absence of superradiance, while the lighter blue region represents the exclusion limits when superradiance is taken into account.
The inclusion of superradiance not only increases the radiation background from axion-like particles but also shifts the parameter space towards regions that would otherwise be excluded in the non-superradiant case. We note that if moduli exist and have a mass that allows superradiance to occur, superradiance must be taken into account when studying PBHs. \\
In order to better appreciate the role superradiance has played so far, it's also instructive to consider what happens when moduli do not exist at all. In this case, not only is superradiance absent, but the only remaining mechanism contributing to $\Delta N_{\rm eff}$ is the radiation of axions through Hawking radiation (we assume that all the ALPs are produced by PBHs and are not present in the primeval radiation). This scenario serves as a useful baseline for comparison, showing the isolated effect of Hawking radiation without the additional contribution of superradiant moduli production.
Hence in Fig.~\ref{neffnomod}, we present the contour lines of $\Delta N_{\rm eff}$, in a scenario where no moduli exist, and therefore, the mechanism of superradiance is entirely absent. The spin of the PBH is set at $a_\star = 0.999$ and at $a_\star = 0$ (Schwarzschild BH).
In contrast to the previous figure, here we observe that the values of $\Delta N_{\rm eff}$ remain relatively low across the parameter space. 
Without the presence of moduli and superradiant amplification, the production of axions is driven solely by Hawking radiation, which is less efficient at increasing the radiation density compared to the combined effects of Hawking radiation and superradiance. 
The value of $\Delta N_{\rm eff}$ can be compared with the value in \cite{Schiavone:2021imu,Bernal:2021bbv}, where similar results were obtained. We observe that rotating PBHs are less efficient than non-rotating black holes for ALP production. This is because $\epsilon_{\rm SM}$ increases with $a_\star$ more than $\epsilon_a$, and thus the ratio $\rho_a/\rho_{\rm SM}$ is lower in the Kerr case compared to the Schwarzschild case.

\section{Conclusions}\label{conclusions}
In this work, we explored the interplay between Hawking radiation and superradiance in the context of Light
Primordial Black Holes (LPBHs) and their implications
for axion-like particles (ALPs) as contributors to Dark
Radiation. We focused on the production of heavy scalar
particles, known as \textit{moduli}, via superradiance instabilities in Kerr black holes and their subsequent decay into
ALPs.\\
We demonstrated that the combined effects of Hawking radiation and superradiance significantly amplify the
production of ALPs, particularly when moduli fields are
present. Using current constraints on the extra effective
number of neutrino species, $\Delta N_{\text{eff}}$, from Planck satellite
observations, we derived updated bounds on the parameter space of this scenario. Our analysis showed that the
inclusion of superradiance extends the range of $\Delta N_{\text{eff}}$ values, potentially shifting the allowed parameter space
compared to scenarios without superradiance. Future experiment CMB-S4 will be able to prove values of $\Delta N_{\text{eff}}$ up to 0.06 \cite{Baumann:2015rya}, thus hinting or further constraining present
scenario. This underscores the importance of superradiance as a mechanism for enhancing the production of
non-standard particles in the early universe.\\
In future work, we plan to further constrain the parameter space by studying the impact of ALPs on reionization or the diffuse X-ray background, resulting from
ALPs reconversion in cosmic magnetic fields or decay into
photons. Moreover, the study can be extended to other
types of Dark Radiation, such as Gravitational Waves or
Dark Photons \cite{Fabbrichesi:2020wbt}.

\section*{Acknowledgments}

This research was partially supported by the research Grant No. 2022E2J4RK ``PANTHEON: Perspectives in Astroparticle and Neutrino THEory with Old and New messengers" under the program PRIN 2022 funded by the Italian Ministero dell’Universit\`a e della Ricerca (MUR).  This work is (partially) supported by ICSC – Centro Nazionale di Ricerca in High Performance Computing.  This article is based upon work from COST Action COSMIC WISPers CA21106, supported by COST (European Cooperation in Science and Technology).

\bibliography{bibliography}

\end{document}